\documentclass[twocolumn,showpacs,preprintnumbers,amsmath,amssymb]{revtex4}
\begin{document}

\title{Comparison of area spectra in loop quantum gravity}
\author{G.~Gour\thanks{E-mail:~gilgour@phys.ualberta.ca} and
V.~Suneeta\thanks{E-mail:~suneeta@phys.ualberta.ca}}
\affiliation{Theoretical Physics Institute,
Department of Physics, University of Alberta,\\
Edmonton, Canada T6G 2J1\\}

\pacs{PACS numbers:}

\begin{abstract}
We compare two area spectra proposed in loop quantum gravity in different 
approaches to compute the entropy of the Schwarzschild black hole. We
describe the black hole in general microcanonical and canonical area
ensembles for these spectra. We show that in the canonical ensemble,
the results for all statistical quantities for any spectrum can be 
reproduced by a heuristic picture of Bekenstein up to second order.
For one of these spectra - the equally-spaced
spectrum - in light of a proposed connection of the black hole 
area spectrum to the quasinormal mode spectrum and following hep-th/0304135,
we present explicit calculations to argue that this spectrum is completely
consistent with this connection. This follows without requiring a
change in the gauge group of the spin degrees of freedom in this formalism
from $SU(2)$ to $SO(3)$. We also show that independent of the area spectrum,
the degeneracy of the area observable is bounded by $C A\exp(A/4)$, where $A$ 
is measured in Planck units and $C$ is a constant of order unity.
\end{abstract}

\maketitle

\section{Introduction}

In recent years, the question of microscopic origin of black hole entropy 
has been addressed in the framework of canonical quantum gravity
\cite{abck,ABK}. A crucial feature of this framework is the fact that operators
corresponding to lengths, areas and volume have discrete spectra \cite{rs}.
A basis for the Hilbert space of canonical quantum gravity is given in 
terms of graphs whose edges carry representations of the group
$SU(2)$. These are known as spin networks as the representations are labeled
by positive half-integers $j$. Each such edge of a spin network contributes
an area $A(j)$ to a surface it intersects, where 
\begin{eqnarray}
A(j) = 8\pi l_{P}^2 \gamma \sqrt{j(j+1)}\;.
\label{areaj}
\end{eqnarray}
The area of a surface $A_S$ is a sum over contributions from all the spin 
network edges intersecting it. Thus, if there are $N$ intersections of 
the surface where each intersecting edge carries spin $j_N$, 
\begin{eqnarray}
A_S = 8\pi l_{P}^2 \gamma \sum \sqrt{j_N (j_N +1)}\;.
\label{lqgs}
\end{eqnarray} 

This spectrum (\ref{lqgs}), which we shall refer to subsequently as the
LQG (Loop Quantum Gravity) spectrum, has been used in \cite{abck} to 
obtain the entropy of a large Schwarzschild black hole. However, the idea of
a discrete area spectrum describing the entropy of a black hole is not new.
The description of black hole entropy using an {\em equally spaced} area 
spectrum has been extensively discussed in the past beginning with heuristic 
arguments \cite{bek2,muk,bek3}. More recently (and more rigorously),
this spectrum has received support from an
algebraic approach to black hole quantization \cite{bek5,bek6,gg2x},
the reduced phase space approach initiated in
\cite{kun1,kun2,gj1}, and from a  WKB treatment
\cite{mak1}.\footnote{More
references  can be found in \cite{bek6}.}  
More recently, a semiclassical description
of the Schwarzschild black hole as a coherent state in LQG has been proposed
by Dasgupta~\cite{dasgupta}; in this description, the area spectrum relevant
for black hole entropy is
\begin{eqnarray}
A_S = 8\pi l_{P}^2 \gamma \sum (j_N + 1/2)\;.
\label{ess}
\end{eqnarray}
We refer to this area spectrum (\ref{ess}) as the ES (Equally Spaced) area
spectrum (this term is used sometimes in literature
for a spectrum without the $1/2$ in the equation (\ref{ess})). 
There seem to be objections to this spectrum in a set of papers by Corichi
and others~\cite{Corichi1,Zapata}. These objections refer to a situation in which the above spectrum
is obtained as the eigenspectrum of an area operator in LQG. One
of the main objections then has to do with the fact that this spectrum predicts
a non-zero value for the area quanta for spin $j=0$. However, in LQG, adding
a spin $j=0$ should not change physics.
We wish to emphasize here
that in~\cite{dasgupta} the ES spectrum is the spectrum of the area of the {\em black hole 
horizon} as measured on a coherent or semiclassical state that describes
the black hole in LQG. Thus, here spin $j=0$ only describes a peaking around
the classical area value. Since this spectrum does not arise out of eigenvalues
of an operator acting on an {\em exact eigenstate}, it is not contradictory
to the arguments in \cite{Corichi1,Zapata} which pertain to a proposal 
in~\cite{alekseev}. In~\cite{alekseev}, Eq.~(\ref{ess}) is claimed  as the correct 
area
spectrum in canonical quantum gravity including quantum corrections. 

One can consider a general area spectrum (which we refer to later as
the GA spectrum) of the form~\footnote{In a
Lorentz covariant loop gravity a completely different
area spectrum has been predicted~\cite{AV}}
\begin{eqnarray}
A_S = 8\pi l_{P}^2 \gamma \sum a(j_N)\;,
\label{gas}
\end{eqnarray}
where $a(j_N)$ is some general function of the spins $j_N$. 

In this paper, we analyze the 
consequences of using various area spectra to describe black hole entropy.
In particular, working in the microcanonical ensemble, we show that for the 
ES area spectrum, the leading order contribution to the entropy comes from
a configuration where nearly all the spins have the value $j=1$. 
This has been stated before in a canonical ensemble language in~\cite{alekseev,poly}. 

This is very 
interesting in light of a proposed connection between the asymptotic
quasinormal mode spectrum and area spectra in canonical gravity by 
Dreyer \cite{dreyer}. We show that applying the argument in~\cite{dreyer} 
carefully to 
the ES spectrum (\ref{ess}) leads to the correct value of semi-classical
entropy {\em without} the need to change the gauge group 
from $SU(2)$ to $SO(3)$. This provides an alternative way of making a 
connection between quasinormal mode and area spectra. This argument for the ES
spectrum was outlined by Polychronakos~\cite{poly}. Here we provide further motivations
and explicit calculations for this argument. There are other ideas that in the context of the
LQG spectrum do not require a change of gauge group from $SU(2)$ to $SO(3)$ for making contact with the QNM spectra~
\cite{xx,swain}. However, as opposed to the explicit computations here, they
are interesting, but heuristic and not by themselves conclusive.

We study the canonical ensembles for these area spectra. The canonical
ensemble in the statistical sense cannot be defined here because there is
no natural notion of a heat bath. However, what we mean by the term is
allowing for fluctuations in the black hole area such that the average of
the area is fixed - then the canonical ensemble maximizes the entropy. 
From this ensemble, we infer that the degeneracy of the area observable, $g(A)$,
is bounded by $C A\exp(A/4)$ for {\em any} area spectrum.   
The most probable configuration of spins
describing a non-Planckian black hole in the ensemble picture is made
up of nearly all equal
spins with value $j= 1/2$ for the LQG spectrum and $j=1$ for the ES spectrum.
Further, we show that in any process leading to a change in area such that one is 
still in the non-Planckian regime, mostly the number of spins with $j= 1/2$ for
the LQG spectrum and $j=1$ for the ES spectrum changes - at leading
order in area. Thus a process leading to a net change in horizon area can be
thought of as occurring due to emission or absorption of spins of mostly 
{\em one} value.

We show that the canonical ensemble in LQG can be compared with the algebraic approach 
initiated by Bekenstein~\cite{bek5,bek6,gg2x} up to second order (in entropy) because the spins other than the most probable
one contribute only at higher orders. This is also the reason why
both in LQG \cite{km} and using the algebraic approach \cite{gg2x}, one obtains the same quantum logarithmic correction $-3/2\ln A$.
One can simulate all the statistical results in LQG within the algebric
approach. i.e. the algebric approach is a very good approximation (up to second order) 
for statistical calculations in LQG. The importance of this argument is manifested
for example when one considers a Schwarzschild BH in a box or equivalently in an $AdS$ spacetime. The statistical 
calculations such as the entropy, area fluctuations etc have been calculated in~\cite{gj2}, 
where the area spectrum and degeneracy of the algebraic approach are used.
Within the spin network approach, when the black hole is in a box (or an $AdS$ spacetime), 
such calculations are extremely complicated and have not been done to the best of our knowledge. 
However, our argument shows that all the calculations performed in~\cite{gj2} for the Schwarzschild black holes 
are the same (up to second order in entropy) as the ones in the spin network approach of LQG.   

\section{Black hole entropy in the microcanonical ensemble}

In computations of black hole entropy in canonical gravity, it is natural
to count microscopic states corresponding to a classical horizon area.
Thus, this is like a microcanonical ensemble
where the {\em horizon area} is fixed. The microscopic states are the spin 
network states that intersect the horizon surface. Each spin $j$ that 
intersects the surface has an internal degeneracy that for large area
black holes is $(2j+1)$. However, arbitrary spins are not allowed - given
a general area spectrum, the spins must satisfy (\ref{gas}). This is the
constraint of fixed area and one counts over all spin network states
corresponding to a given classical horizon area. Thus, the degeneracy of
states corresponding to a given area $A$, denoted by $g(A)$ is a sum over
contributions $g(\{j_N\})$ 
\begin{eqnarray}
g(\{j_N\}) \approx \prod_{\{j_N\}} (2j_N +1)
\label{deg}
\end{eqnarray}
and each set $\{j_N\}$ satisfies the area constraint (\ref{gas}). 
The
exact expression for the degeneracy corresponding to a set of spins is
given by the dimension of the space of conformal blocks of an
$SU(2)_{k}$ WZW conformal field theory, see for e.g \cite{km}. In this
theory, for large area, this is approximately equal to (\ref{deg})
above.
For a 
general area spectrum, the degeneracy $g(A) = \sum_{\{j_N\}} g(\{j_N\})$. 

This ensemble must be carefully defined for the LQG spectrum (\ref{lqgs}).
In general, when we try to define a microcanonical area ensemble - i.e
counting the quantum states corresponding to a given classical area - 
we need to define an appropriate spread around the area since the area 
spectrum is discrete.
The constraint (\ref{lqgs}) would be satisfied
in general only for some choices of spin network configurations.
This may cause
the entropy to change sharply as one goes from one classical area to another.
For example, choosing a (large) value
of $A$ such that it is saturated by $N$ spins of $j=1/2$, we see that 
$N \sim A/l_{P}^2$ and $S = \ln g(A) \sim A$. On the other hand, one could
choose a large $A$ with a value such that it cannot be saturated by spins
with $j=1/2$.
In this case, the entropy could be different - either proportional to area with
a different proportionality constant, or not even proportional to the area.
However, for this area
spectrum to describe black hole entropy, one must get an entropy 
proportional to area with a fixed proportionality constant. 
The solution is to choose
a spread large enough so that it contains a value 
of the area that is saturated by spins of $j =1/2$. At
first sight, this appears contrived. But due to the special 
properties of this spectrum \cite{steve}, there is a very large
number of states lying even in a Planckian size spread in area, as long
as the area is large. In fact, for a large horizon area $A_H$, 
the number of area eigenvalues in the range $A_{H}\pm l_{P}^{2}$ is of the order of
$\exp(\sqrt{A_{H}}/l_{p})$ (The number of area eigenvalues is not to be 
confused with the number of configurations corresponding to a fixed classical
area, which gives the main contribution to the entropy).
 
Therefore, in \cite{abck} where
the Schwarzschild black hole entropy is computed using the LQG spectrum,
it is argued that one must `trace over' states lying in an area range
$A \pm l_{P}^{2}$. The counting of states corresponding to this spread is
done there - the final result is that the contribution from spins with
$j=1/2$ gives an entropy proportional to area, and the rest of the spins
give a lesser contribution each - however the sum is not evaluated 
explicitly. 
We work later
in a canonical ensemble allowing not for Planckian, but arbitrary
fluctuations keeping the average of area fixed, and obtain a {\em bound} 
on this sum. The sum of the
contributions from other spins to the entropy is bounded by
the logarithm of the area.
This `logarithmic' correction to the entropy which comes from the other spins 
and area fluctuations is distinct from the quantum corrections discussed in
\cite{km}.

We now study the ES spectrum in this picture. From the form of the area
constraint (\ref{ess}), we see that for a given classical area, we 
must choose an
appropriate spread around the area such that we have an area value $A$ in the
spread for which
$A/(8\pi \gamma l_{P}^{2})$
is a half-integer $Q$. Nothing more is required, and there are now
many possible sets of spins $\{j_N\}$ - corresponding to the number
of partitions of $2Q$. We first consider the case when all
spins corresponding to a particular realization of area are equal. In this case,
the area constraint is 
\begin{eqnarray}
A = 8\pi \gamma l_{P}^{2} N_{j} (j + 1/2)
\label{eqs}
\end{eqnarray}
where $N_j$ is the number of spins of value $j$ required to satisfy the area
constraint. Then the degeneracy corresponding to each such set can be
written as $g(\{j_N \}) = g(j, N_j )$ and
\begin{eqnarray}
g(j, N_j) = (2 j + 1)^{N_{j}}
\label{degeqs}
\end{eqnarray}
For large areas, $N_j$ is large (of $O(Q)$) when $j$ is small. Thus the 
degeneracy is more for small spin values. We however see something surprising.
Let us look at the degeneracies for the cases when all the spins have $j=1/2$,
$j=1 $ and $j=3/2$. We have
\begin{eqnarray}
&\;& g(1/2, N_{1/2}) = 2^Q \; ;\;\; g(1, N_{1}) = (3^{2/3})^Q \nonumber\\ 
&\;& g(3/2, N_{3/2}) = 2^Q
\label{gvalues}
\end{eqnarray}
Thus it is easy to see that the degeneracy is maximum when we saturate
the area with $j=1$, and
furthermore, $j=1/2$ and $j=3/2$ have the same degeneracies. Higher spins 
will have smaller degeneracies. We have only considered the case of equal
spins so far. However, it will be shown in the following that any combination 
of unequal spins yields a lesser contribution than the case with all spins $j=1$ 
for large areas. 

In the general case we have $N_{1}$ spins with $j=1$ and $N_{j_{i}}$ spins equal to $j_{i}$
for each $i=2,3,...,s$. The area constraint can be written simply as
\begin{equation}
Q={3\over 2}N_1+\left(j_{2}+{1\over 2}\right)N_{j_{2}}+...
+\left(j_{s}+{1\over 2}\right)N_{j_{s}}\;,
\end{equation}
and the degeneracy is
\begin{eqnarray}
g & = & 3^{N_{1}}(2j_{2}+1)^{N_{j_{2}}}\cdots (2j_{s}+1)^{N_{j_{s}}}\nonumber\\
& = & \left(3^{2/3}\right)^{Q}\left(\frac{2j_{2}+1}{3^{(2j_{2}+1)/3}}\right)^{N_{j_{2}}}
\cdots \left(\frac{2j_{s}+1}{3^{(2j_{s}+1)/3}}\right)^{N_{j_{s}}}  
\end{eqnarray}
Now, for $j_{i}\neq 1$ it easy to check that $(2j_{i}+1)/3^{(2j_{i}+1)/3}<1$, 
and therefore $g<(3^{2/3})^{Q}$.
Since the degeneracy corresponding to all spins having $j=1$ is
$g(1, N_1) = (3^{2/3})^Q$, we see that any such combination of spins as
above always yields a lesser degeneracy.
Hence, for large area $A$, we can write the degeneracy as 
\begin{eqnarray}
g(A) \approx (3^{2/3})^Q\;, 
\label{degess}
\end{eqnarray}
and the entropy is 
\begin{eqnarray}
S = \ln g(A) = \frac{\ln 3}{12 \pi \gamma l_{P}^{2}}~A +...
\label{essentropy}
\end{eqnarray} 

Clearly, for large areas, the leading configuration contributing to
the entropy will always be equal spins with $j=1$. 
The above result for the entropy (the appearance of a factor of $\ln 3$)
is also very striking in that the choice of
an ES spectrum has naturally led to a dominance of configurations with
spin $j=1$. Recently, there has been much excitement over a possible 
connection between the asymptotic quasinormal mode spectrum of the black
hole and the area spectrum. In~\cite{dreyer} this connection is used to
fix the value of the Immirzi parameter $\gamma$ and it is shown that for the
LQG spectrum it gives the correct semi-classical entropy. However, 
in~\cite{dreyer}, it is claimed that using the same argument,
the ES spectrum would not yield the correct semi-classical entropy. We
now proceed to show that this is not the case; applying the argument
carefully, one indeed obtains the correct semiclassical entropy. There is
the added advantage that with the ES spectrum, the factor of $\ln 3$ needed
to make contact with quasinormal mode spectra appears naturally in the 
entropy - {\em without} the need to argue for a change of gauge group
from $SU(2)$ to $SO(3)$ as required for the LQG spectrum. Both the observation
that spins $j=1$ dominate for the ES spectrum and  that 
making a connection to the quasinormal mode spectrum yields the correct
semi-classical entropy has been made before in a canonical ensemble
picture in \cite{alekseev, poly}. We present explicit calculations that 
are necessary to motivate this argument.
We apply the argument in~\cite{dreyer} to a general area spectrum and then
specialize to the case of the ES spectrum. First we revisit the
argument which is motivated by the following result : 

The reaction of 
a black hole to a perturbation is given for certain intermediate times
by a set of damped oscillations which are characteristic of the black hole.
These are the quasinormal modes (See~\cite{nollert,kokkotas} for a review).
It is an interesting result that for large imaginary part of the quasinormal
frequency (higher order modes), the asymptotic quasinormal mode spectrum 
for a Schwarzschild black hole can be computed analytically 
\cite{motl,neitzke} for scalar or gravitational perturbations. 
It is
\begin{eqnarray}
M\omega = \frac{1}{8\pi}~ \ln 3 + \frac{i}{4} \left(n - {1\over 2}\right)
\label{asyqnm}
\end{eqnarray} 
This behavior was found approximately from numerical calculations 
first in \cite{nollert1}
(for Kerr black holes, the highly damped QNMs are computed numerically in~\cite{cardoso}) and it was recognized by Hod \cite{hod} that the 
numerical equation in \cite{nollert1} resembled (\ref{asyqnm}). 
Hod then suggested (based on an argument of Bekenstein~\cite{bek}) that in 
the spirit of Bohr's correspondence 
principle, an oscillatory frequency describing the black hole's (classical)
response to a perturbation (i.e real part of the quasinormal frequency) was
related to a transition frequency in an area spectrum describing a change
in horizon area $\Delta A$. 

The exact relation was the following:
The real part of the quasinormal mode spectrum (\ref{asyqnm}) is given 
by 
\begin{eqnarray}
\omega_{QNM} = \frac{\ln 3}{8\pi M} 
\label{omegaqnm}
\end{eqnarray}
Then the change in mass of the black hole $\Delta M$ due to a 
transition is equal to $\hbar \omega_{QNM}$. Since for a Schwarzschild black hole,
horizon area $A = 16 \pi M^2$, this implies that 
\begin{eqnarray}
\Delta A = 4 l_{P}^{2}\ln 3
\label{deltaa}
\end{eqnarray} 

In \cite{dreyer}, the change in area $\Delta A$ is computed for the LQG
spectrum and equated to (\ref{deltaa}); this fixes the value of the 
Immirzi parameter. This value gives the correct semi-classical entropy
if one considers $SO(3)$ spins instead of $SU(2)$ spins. This conclusion is
correct for the LQG spectrum. However, there is an important point to note
when one considers a different area spectrum. In the framework of spin 
network degrees of freedom intersecting a surface, a change of area of the
surface is caused by an emission or absorption of a puncture with spin. 
When this surface is the black hole horizon, we have already seen that 
the configuration (of set of punctures with spins) that dominates the
entropy depends on the choice of spectrum. When the spectrum is such
that this configuration is made up of equal spins $j_{MP}$ (MP stands for
Most Probable configuration), it is clear that the {\em minimum} change
in horizon area $\Delta A$ would arise from the absorption or emission
of a spin in the most probable configuration (We provide a proof of this
statement in the canonical ensemble picture in the next section). 
Thus given a general area
spectrum of the form (\ref{gas}), 

\begin{eqnarray}
\Delta A = 8\pi \gamma l_{P}^{2} a_{j_{MP}}
\label{deltaagas}
\end{eqnarray}

For the LQG spectrum 
it is indeed correct that $j_{MP} = j_{min}$ where
$j_{min}$ is the minimum value of spin for the gauge group chosen, either
$SU(2)$ or $SO(3)$. Then, as argued in~\cite{dreyer}, on fixing the value of the 
Immirzi parameter using the quasinormal mode spectrum, the correct 
semiclassical entropy is obtained by choosing a gauge group $SO(3)$.
For the ES spectrum with a choice of gauge group $SU(2)$, as we saw,
$j_{MP} = 1$ and $j_{min} = 1/2$. 

Let us now fix the value of the Immirzi parameter for a general area spectrum
of the form (\ref{gas}). We thus equate the r.h.s of (\ref{deltaa}) and
(\ref{deltaagas}). Then we get
\begin{eqnarray}
\gamma_{GA} = \frac{\ln 3}{2\pi a_{j_{MP}}}
\label{gammagas}
\end{eqnarray}

For the ES spectrum, $j_{MP} = 1$ and $a_{j_{MP}} = 3/2$. Thus for this
spectrum, 
\begin{eqnarray}
\gamma_{ES} =  \frac{\ln 3}{3\pi}. 
\label{gammaes}
\end{eqnarray}
Substituting this value of the
Immirzi parameter into the expression for entropy with the ES spectrum
(\ref{essentropy}), we see that we recover the correct
expression for semi-classical entropy. In fact, for a general area spectrum
with the Immirzi parameter (\ref{gammagas}), as long as its maximum
entropic configuration is given in terms of equal spins $j_{MP}$ - we
have the correct semi-classical entropy if $j_{MP} = 1$. Dreyer's argument for
replacing $SU(2)$ by $SO(3)$ as the gauge group is a particular case of
this general statement.

\section{Black hole entropy in the canonical ensemble}

There are, in fact, {\it two} distinct (but not always separable) sources for 
black hole entropy. Firstly, there should be a contribution to the entropy due to 
the number of microstates that are necessary to describe a black hole of a fixed 
horizon area. Secondly, since in dynamical processes, a black hole absorbs
or emits radiation/matter across the horizon, the horizon area will fluctuate.  
This leads to an additional contribution to the entropy.

It therefore is of interest to study the black hole in an area ensemble that
allows for fluctuations in area such that the average of the area is a 
fixed `classical' value - the term used to denote the situation when area 
fluctuations are caused by `quantum processes' (i.e absorption or emission
of spins). This is the canonical area ensemble - however unlike the usual
canonical ensemble in statistical mechanics, there is no natural analogue
of a heat bath here (There is such an analogue for black holes in $AdS$ 
spacetimes, as discussed in \cite{gj2}). 
 
In a quantum description, black holes can be described by a density matrix, $\rho_{bh}$.
In particular, in canonical gravity, the density matrix relevant for 
black hole entropy is obtained by tracing over the `volume' states (the 
total Hilbert space carries information about both the bulk and the surface degrees of 
freedom). The entropy of the black hole
is given by 
\begin{equation}
S=-{\rm Tr}\rho_{bh}\ln\rho _{bh}
\label{entropy}
\end{equation}
where $\rho_{bh}$ is the density matrix describing a black hole
with an average area
\begin{equation}
A_{H}={\rm Tr}{\bf A}\rho _{bh},
\label{con1}
\end{equation}
where ${\bf A}$ is the area operator whose eigenvalues give the area spectrum. 

Our strategy is to find a density matrix that maximizes the entropy 
given in (\ref{entropy}) and satisfies the condition (\ref{con1}). From a physical
point of view, this is {\it not} the only requirement on the density matrix. In order to 
achieve thermal equilibrium one has to place the black hole
in a `box' (or, for example, in an AdS spacetime). This will induce more conditions,
and in a first order approximation one expects a Gaussian  
distribution around the mean 
value $A_{H}$. That is, the introduction of the box imposes some particular value, $\Delta$ 
on the fluctuations of the area (which depends also on the parameters of the box): 
\begin{equation}
\Delta ^{2} =\langle {\bf A}^{2}\rangle -\langle {\bf A}\rangle ^{2}
={\rm Tr}{\bf A}^{2}\rho _{bh} - A_{H}^{2}
\label{con2}
\end{equation} 
Therefore, the density matrix, ${\bf \rho}_{bh}$, that maximizes the entropy 
given in (\ref{entropy}) and satisfies both conditions (\ref{con1}) 
and (\ref{con2}) is given by:
\begin{equation}
\rho _{bh}=\frac{1}{Z}\exp(\eta {\bf A}-\mu {\bf A}^{2}) 
\label{restrho}
\end{equation}
where $\eta$ and $\mu$ are Lagrange multipliers and the partition function $Z$ is 
determined by the requirement ${\rm Tr}\rho_{bh} =1$.
However, as we stated earlier, there is no natural analogue of a 
physical heat bath 
here, and no mechanism by which fluctuations would be restricted as in
(\ref{con2}).  

In the 
computation of black hole entropy using the LQG spectrum \cite{abck}, classically 
there is no radiation across the horizon. 
However, quantum fluctuations of the classical area $A \pm l_{P}^{2}$ are
considered. In fact, they are necessary to get a result $S \sim A$. The 
physical picture is probably one where arbitrary area fluctuations are not 
allowed, but quantum fluctuations of the order of Planckian area, caused
by spin emissions across the 
horizon are. This can be stated in the canonical ensemble picture as a
condition of the form (\ref{con2}) where the fluctuation
$\Delta \sim O(l_P )$. Thus, one must use the density matrix (\ref{restrho})
where $\Delta \sim O(l_P )$ would correspond to a particular regime for the
Lagrange multipliers $\eta$ and $\mu$. This unfortunately is a very
complicated exercise for the LQG spectrum. We are forced to ignore 
(\ref{con2}) and allow arbitrary fluctuations keeping the area average
fixed. Since we now place lesser constraints on the system, the entropy
will be {\em more} - 
and will provide an {\it upper bound} on the 
actual entropy using the LQG spectrum with Planck area quantum fluctuations. 

For arbitrary fluctuations keeping the area average fixed, the 
relevant density matrix  $\rho _{m}$ (for a general area spectrum)
is given according to 
standard statistical mechanics by
\begin{equation}
\rho _{m}=\frac{1}{Z}\exp(-\lambda {\bf A}),
\label{dmatrix}
\end{equation}
where the Lagrange multiplier, $\lambda$ will be calculated below, and
\begin{equation}
Z={\rm Tr}\exp(-\lambda {\bf A}).
\end{equation}
Substituting $\rho _{m}$ in (\ref{entropy}) we find
\begin{equation}
S_{m}\equiv -{\rm Tr}\rho _{m}\log\rho _{m}=\ln Z + \lambda A_{H}
\label{entm}
\end{equation}
(note that $A_{H}\equiv \langle {\bf A}\rangle = {\rm Tr}{\bf A}\rho _{m}$).

Following \cite{alekseev,poly}, we shall now compute the partition function in this spin network picture for
a general area spectrum of the form (\ref{gas}). Then we shall specialize to
the cases of the LQG and ES spectra.
The partition function $Z(\lambda)$ is given by:
\begin{equation}
Z(\lambda)=\sum _{\{j_{N}\}}\prod_{j_{i}\in\{j_{N}\}}(2j_{i}+1)\exp\left[-\lambda A(\{j_{N}\})\right],
\end{equation}
where $A(\{j_{N}\})$ for the general case is given in (\ref{gas}) and 
the product term is approximately the degeneracy for the given set $\{j_{N}\}$ for large horizon area 
black holes (The exact expression for the degeneracy can
be found in \cite{km}).
Let us now denote by $m_{i}$ ($i=1,2,...$) the number of times the
value $i/2$ appears in the set $\{j_{N}\}$. In this notation the area
can be written as:
\begin{equation}
A(\{j_{N}\})=8\pi l_{p}^{2}\gamma\sum_{i=1}^{\infty}m_{i}a_{i}\;,
\end{equation}
where $a_{i}\equiv a(j=i/2)$. Thus, in this notation, the partition
function has the form 
\begin{equation}
Z(\lambda)=\sum_{\{m_{i}\}}\prod_{i=1}^{\infty}(i+1)^{m_{i}}\exp(-8\pi
l_{p}^{2}\gamma\lambda m_{i}a_{i})\;,
\label{partit}
\end{equation}
where the sum over all the sets $\{j_{N}\}$ has been replaced by the
sum over all the sequences $\{m_{i}\}$ of integers.

The partition function above converges if and only if for all
$i=1,2,...$ we have $(i+1)\exp(-8\pi l_{p}^{2}\gamma\lambda a_{i})<1$.
This condition imposes a lower bound on $\lambda$; i.e. the partition
function $Z(\lambda)$ converges if and only if $\lambda >\lambda _{c}$, where
\begin{equation}
\lambda _{c}\equiv\lambda _{i_{c}}\equiv {\rm max} \{\lambda _{i}\}\;,
\end{equation}
and $\lambda _{i}$ is defined by the requirement\\ 
$(i+1)\exp(-8\pi l_{p}^{2}\gamma\lambda_{i} a_{i})=1$, i.e.
\begin{equation}
\lambda_{i}=\frac{\ln(i+1)}{8\pi l_{p}^{2}\gamma a_{i}}
\end{equation}
Thus, for $\lambda >\lambda _{c}$
\begin{equation}
Z(\lambda)=\prod_{i=1}^{\infty}\frac{1}{1-(i+1)\exp(-8\pi l_{p}^{2}\gamma\lambda a_{i})}\;.
\label{part}
\end{equation}
For the LQG spectrum, $a_{i}=\sqrt{\frac{i}{2}(\frac{i}{2}+1)}$, and 
$\lambda_{c}=\lambda_{1}$, i.e corresponding to a spin $j = 1/2$.
In comparison, for the ES spectrum, $a_{i}=\frac{i+1}{2}$, and
in this case $\lambda_{c}=\lambda_{2}$, corresponding to a spin $j = 1$.

The average of the area is given by
\begin{eqnarray}
A_{H} & \equiv & {\rm Tr}\rho _{m}{\bf A}=-\frac{d}{d\lambda}\ln Z \nonumber\\
& =& 
8\pi l_{p}^{2}\gamma\sum_{i=1}^{\infty}\frac{(i+1)a_{i}\exp(-8\pi l_{p}^{2}
\gamma\lambda a_{i})}{1-(i+1)\exp(-8\pi l_{p}^{2}\gamma\lambda a_{i})}.
\label{area}
\end{eqnarray}
Now, since we are interested in large black holes 
(compared to the Planck scale), 
$A_{H}/l_{p}^{2}\gg 1$. However, in the above equation, the term in the numerator
$(i+1)a_{i}\exp(-8\pi l_{p}^{2}\gamma\lambda a_{i})$ approaches zero very
fast as $i$ increases when $\lambda >\lambda_{c}$. Thus, the only way to get
large $A_{H}$ is by the requirement that one of the terms in the
denominator in (\ref{area}) be close to zero. Fortunately,
this is possible when $\lambda\rightarrow\lambda_{c}=\lambda_{i_{c}}$. This
picks up as the {\em leading} term in (\ref{area}), 
the one corresponding to spins with value $i_{c}/2$.
Therefore,
the partition function for $A_{H}/l_{p}^{2}\gg 1$ (or equivalently
$\lambda \approx \lambda_c $) is given by:
\begin{equation}
Z(\lambda)\approx \frac{1}{1-(i_{c}+1)\exp(-8\pi l_{p}^{2}\gamma\lambda a_{i_{c}})}.
\label{largeaz}
\end{equation}
In order to obtain the relation between $A_{H}$ and $\lambda$,
we substitute $\lambda=\lambda_{c}+\delta$ in Eq.~(\ref{area}) and
expand it up to the second order in $\delta$. We find that
\begin{equation}
A_{H}=\frac{1}{\delta}[1+(c-4\pi l_{p}^{2}\gamma a_{i_{c}})\delta +O(\delta ^{2})]\;,
\end{equation} 
where the constant $c$ is of the order of Planck area and is given by
\begin{equation}
c=\sum_{i\neq i_{c}}\frac{a_{i}(i+1)\exp(-8\pi l_{p}^{2}
\gamma\lambda_{c} a_{i})}{1-(i+1)\exp(-8\pi l_{p}^{2}
\gamma\lambda _{c} a_{i})}\;.
\end{equation}
Hence, the relation between $A_{H}$ and $\lambda$ is given by:
\begin{equation}
\lambda=\lambda_{c}+\frac{1}{A_{H}}+O(l_{p}^{2}/A_{H}^{2})\;.
\label{largeal}
\end{equation}
Substituting this in (\ref{entm}) gives
\begin{equation}
S_{m}=\lambda_{c}A_{H}+\ln A_{H} +O(1)
\label{canentropy}
\end{equation}
This is an {\it upper} bound on the entropy of the black hole defined in 
(\ref{entropy}) for a general area spectrum (\ref{gas}) in
the spin network picture. In particular, this is also an upper bound for the entropy of a density matrix that satisfies
the average area constraint (\ref{con1}) and has zero area fluctuations, i.e. the micro-canonical ensemble. 
Therefore, the degeneracy 
\begin{equation}
g(A)\leq\exp(S_{m})=C{A_{H}\over\l_{P}^{2}}\exp\left(\frac{A_{H}}{4 l_{P}^{2}}\right)\;,
\end{equation}
where $C$ is a constant of order unity.
Before seeing the consequences for the LQG and ES
spectra in particular, we note something very interesting that is revealed 
by working in the canonical ensemble. 
We saw that for any general area spectrum (\ref{gas}),
the large area limit for the black hole is dominated by the configuration
of equal spins with values $i_{c}/2$.  
In the large area regime, $\lambda$ is given by (\ref{largeal}). Now, if
in a process, the horizon area changes, as long as it is in the large
area regime, it would always be dominated (at {\em leading} order) by a 
similar configuration of equal spins with values $i_{c}/2$ - except that the
{\em number} of these spins would be different. The terms corresponding to
other spin values in (\ref{area}) do not change at leading order! 
Summarizing the results, in the spin network
picture, convergence of the partition function in the canonical ensemble 
implies that :

\begin{itemize}
\item
A large area black hole is described by a most probable 
configuration involving nearly all equal
spins of
{\em one} value, $j = i_{c}/2$. 
\item
In any process leading to a change in horizon area (for e.g, exchange of 
spins across the horizon) such that one is always in the large black
hole regime, on the average, the number of the spins $j = i_{c}/2$ changes at
leading order - the numbers
of all other spins remaining nearly the same. 
We will compute the average number
of spins of a given value in a configuration in this ensemble below to 
demonstrate this point.

\end{itemize}

These results are independent of the particular choice of area spectrum.
Even if the area spectrum we started out with was not equally spaced (i.e
LQG spectrum), area increments are most probably due to exchange of a 
certain number of spins $j = i_{c}/2$. Therefore they are in multiples of a fixed number - so
long as we are in the large area regime. The peculiarity of the spectrum 
shows up only for Planckian black holes.

Let us now illustrate this by calculating the average number of spins with 
$j=j_{c}=i_{c}/2$
and the average number of spins with $j\neq j_{c}$. For this 
purpose, we define an operator ${\bf N}_{j}$ that counts the number
of spins with the value $j = i/2$ for a given configuration of spins.
Thus, the average number of spins with a value $j=i/2$ is
\begin{equation}
\langle {\bf N}_{j} \rangle={\rm Tr}{\bf N}_{j}\rho_{m}
=\sum_{m_{i}}m_{i}(i+1)^{m_{i}}\exp(-8\pi
l_{p}^{2}\gamma\lambda m_{i}a_{i})\;,
\end{equation}
where we have used the same notation as in (\ref{partit}). 
Evaluating the sum above, we find
\begin{equation}
\langle {\bf N}_{j} \rangle =\frac{(i+1)\exp(-8\pi l_{p}^{2}
\gamma\lambda a_{i})}{1-(i+1)\exp(-8\pi l_{p}^{2}\gamma\lambda a_{i})}\;.
\end{equation}
Hence, by substituting the value of $\lambda$ from (\ref{largeal}) we
find that the average number of spins that are not equal to the
critical value is given by
\begin{eqnarray}
N_{j\neq j_{c}} & = &\sum_{j\neq j_{c}}\langle {\bf N}_{j} \rangle\nonumber\\
& = &\sum_{i\neq i_{c}}\frac{(i+1)\exp(-8\pi l_{p}^{2}
\gamma\lambda_{c} a_{i})}{1-(i+1)\exp(-8\pi l_{p}^{2}
\gamma\lambda _{c} a_{i})}
 +  O(1/A) \nonumber\\
& \equiv & c+O(1/A)\;,
\end{eqnarray} 
where $c$ is a constant number (i.e. not depending on $A_{H}$) of
the order of unity. On the other hand, $N_{j=j_{c}}\sim
A_{H}/l_{p}^{2}$.
The fact that for large area $N_{j\neq j_{c}}$ is a constant of order unity
and is independent (at first order) of the value of $A_{H}$ shows
that it is the critical spin that play the major role in any dynamical 
process. 

Consider a black hole with an initial area average $A_{initial}$ 
(we consider macroscopic black holes with area much higher than the
Planck scale). Then, after some dynamical process has occurred -
for example some particles (or in this picture, spins) have been absorbed 
or emitted by the black
hole - the black hole area average changes to $A_{final}$. Since the
average of spins with $j\neq j_{c}$ has not changed in the process, we
can safely conclude that the change in the area occurred mostly due to 
exchange in the number of spins with $j=j_{c}$. Therefore, we conclude
that in any dynamical process the probability to exchange a spin with $j\neq
j_{c}$ is much lower than the probability to exchange a spin with 
$j=j_{c}$. This explains why it is the $j_{MP}\equiv j_{c}$ that
should be taken in the comparison with the quasinormal mode spectrum (and not
$j_{\rm min}$).

In the above discussion we ignored the fact that in our ensemble 
the fluctuations in the area are very large. They are given by
\begin{equation}
\Delta^{2}\equiv\langle {\bf A}^{2}\rangle 
-\langle {\bf A}\rangle ^{2}
=\frac{d^{2}}{d\lambda ^{2}}\ln Z\sim A_{H}^{2}.
\end{equation}
Thus, $\Delta\sim A_{H}$ as it was expected since the black hole is
not in a box and, therefore, cannot be in an equilibrium. However,
putting the black hole in a box will not change the results substantially. The
argument is that in the limit of a very large box the black hole can
be described by $\rho _{m}$. Now, taking a smaller box will only
reduce the fluctuations and therefore will not increase (actually,
probably decrease) the probability to emit/absorb spins. That is,
the probability to emit/absorb a spin with $j=j_{c}$ will remain much 
higher than the probability to exchange a spin with $j\neq
j_{c}$. 

In the algebraic approach the area spectrum is equally spaced, with eigenvalues
$A_{n}=a_{0}n$, where $n$ is an integer and $a_{0}$ is a fundamental unit of area
of the order of Planck scale. The degeneracy of the area levels is approximately $k^{n}$
(there are corrections due to zero hyperspin~\cite{gg2x}), where
$k$ is an integer number greater than 1. The integer $k$ is usually taken to be equal
to 2 or 3. This picture is identical to the spin network approach if one ignores all other spins
but the most probable one. For $k=2$, $j_{MP}=1/2$ and for $k=3$, $j_{MP}=1$.
Following the same steps that led to (\ref{canentropy}), the algebraic approach 
gives the same results for the entropy up to the second order 
(i.e. with exactly the same logarithmic correction). Our argument above shows why -  
the spins other than the most probable one in LQG contribute only at higher orders, even
for different types of density matrices (such as, for example, the density matrix given in 
(\ref{restrho})).  
 
In~\cite{gj2} it has been shown (within the algebraic approach) that 
for an AdS-Schwarzschild black hole (which is in a sense equivalent to 
putting the Schwarzschild black hole in a box due to the nature of the
AdS potential) the correction to the entropy is $1/2\ln A$ and the fluctuations 
in the area are $\Delta\sim A^{1/2}$. Our argument implies that we will get exactly
the same results in the spin network approach (although the calculations in the spin
network approach suffer from mathematical complications that make them difficult to compute). 
Note that the presence of the box only reduces the area fluctuations and the logarithmic 
corrections to the black hole entropy. 

Let us now compute the entropy
for the ES and LQG spectra in the area canonical ensemble.
For the ES spectrum,
the entropy given by (\ref{canentropy}) is :
\begin{eqnarray}
S = A_{H} (\ln 3)/(12\pi l_{P}^{2} \gamma_{ES})  + \ln A_{H} + O(1)
\label{canesentropy}
\end{eqnarray}

Choosing a value for the Immirzi parameter $\gamma_{ES}$ in 
(\ref{gammaes}) obtained by making contact with the quasinormal
mode spectrum, we get the correct semi-classical entropy. The next order
correction arises from area fluctuations in the area canonical ensemble. 

For the LQG spectrum, the entropy from (\ref{canentropy}) is now

\begin{eqnarray}
S = A_{H} (\ln 2)/(4\pi \sqrt{3}l_{P}^{2} \gamma_{LQG})  + \ln A_{H} + O(1)
\label{canlqgentropy}
\end{eqnarray}

One can choose an appropriate value for $\gamma_{LQG}$ such that the 
correct semi-classical entropy is recovered - as done in \cite{abck}.
As shown in \cite{dreyer}, this is not compatible with the Immirzi parameter
obtained from the quasinormal mode spectrum. To make them compatible
requires a change of gauge group to $SO(3)$. The expression 
(\ref{canlqgentropy}) is the entropy for the LQG spectrum with large
area fluctuations keeping the average of area fixed. However,  
in the picture of the quantum black hole in \cite{abck}, no exchange of
radiation is allowed across the horizon except an  exchange of
spins corresponding to Planckian fluctuations in area, i.e  
$A_{H} \pm l_{P}^{2}$. The entropy with fluctuations restricted
to be Planckian is less than or equal to the entropy given by
(\ref{canlqgentropy}) above. In particular, it allows us to add a statement
to the result of \cite{abck} - that the leading order contribution to the 
entropy comes from the spins with least value, and
the sum of contributions from all other spins is bounded by $\ln A$.  

\section*{Acknowledgments} 
We are indebted to Don Page and Valeri Frolov for very valuable discussions,
and to Jacob Bekenstein for his suggestions.
We would also like to thank Arundhati Dasgupta, Fotini Markopoulou, Lee Smolin, 
Thomas Thiemann and Andrei Zelnikov for useful comments and Bob Teshima for help with 
computational programs. 
We would like to thank Alexios Polychronakos for drawing our attention 
to a section in \cite{poly} where the dominance of spins $j=1$ in the ES spectrum
was discussed first. We are grateful to J. Engle for pointing out a missing numerical
factor in one of the equations.	
GG is grateful to the Killam Trust for its financial support. 
VS is supported by a fellowship from the Pacific Institute for the 
Mathematical Sciences and the NSERC of Canada.




\begin{thebibliography}{100}

\bibitem{abck} A. Ashtekar, J. Baez, A. Corichi and K. Krasnov, 
Phys. Rev. Lett. {\bf 80} 904 (1998).

\bibitem{ABK} A. Ashtekar, J. Baez and K. Krasnov,
Adv.\ Theor.\ Math.\ Phys.\  {\bf 4}, 1 (2000)
[arXiv:gr-qc/0005126].

\bibitem{rs} C. Rovelli and L. Smolin, Nucl. Phys. {\bf B42}, 593 (1995).

\bibitem{bek2} J.D. Bekenstein, Lett. Nuovo Cimento, {\bf 11},
467 (1974).

\bibitem{muk} V.F. Mukhanov, JETP Letters {\bf 44}, 63 (1986).

\bibitem{bek3} J.D. Bekenstein and V.F. Mukhanov, Phys. Lett. {\bf B360}, 7
(1995) [gr-qc/9505012].

\bibitem{bek5} J.D. Bekenstein, The case of discrete energy levels of a 
black hole 2001: A Spacetime Odyssey ed M J Duff and J T Liu 
(Singapore: World Scientific) p 21 [hep-th/0107045].

\bibitem{bek6} J.D. Bekenstein and G. Gour, Phys. Rev.
{\bf D66}, 024005 (2002) [gr-qc/0202034].

\bibitem{gg2x} G. Gour, Phys. Rev. {\bf D66}, 104022 (2002)
 [gr-qc/0210024].

\bibitem{kun1} A. Barvinsky and G. Kunstatter, ``Mass Spectrum
for Black Holes in Generic 2-D Dilaton Grvaity'', gr-qc/9607030
(1996).

\bibitem{kun2}  A. Barvinsky, S. Das and G. Kunstatter, 
Class. Quant. Grav. {\bf 18} 4845 (2001)
[gr-qc/0012066].  

\bibitem{gj1} G. Gour and A.J.M. Medved, Class. Quant. Grav. {\bf 20} 2261 (2003) 
[gr-qc/0211089]; Class. Quant. Grav. {\bf 20} 1661 (2003) [gr-qc/0212021].

\bibitem{mak1} J. Louko and J. Makela, Phys. Rev. {\bf D54},
4982 (1996) [gr-qc/9605058];
 J. Makela and P. Repo,
Phys. Rev. {\bf D57}, 4899 (1998) [gr-qc/9708029];
J. Makela, P. Repo, M. Luomajoki and J. Piilonen,
Phys. Rev. {\bf D64}, 024018 (2001) [gr-qc/0012005].

\bibitem{dasgupta} A. Dasgupta, JCAP 0308 004 (2003); arXiv : hep-th/0310069.

\bibitem{Corichi1}  A.~Corichi, gr-qc/0402064.

\bibitem{Zapata} J.~A.~Zapata, gr-qc/0401109.

\bibitem{alekseev} A. Alekseev, A. P. Polychronakos and M. Smedback,
Phys. Lett. {\bf B574} 296 (2003).

\bibitem{AV}
S. Alexandrov, D. Vassilevich, Phys. Rev. D{\bf 64}, 044023 (2001)
[gr-qc/0103105].

\bibitem{poly} A. P. Polychronakos, arXiv : hep-th/0304135.

\bibitem{dreyer} O. Dreyer, Phys. Rev. Lett. {\bf 90} 081301 (2003).

\bibitem{xx} Alejandro Corichi, Phys. Rev. D{\bf 67}, 087502 (2003);
Yi Ling and Hongbao Zhang, Phys. Rev. D{\bf 68}, 101501 (2003).

\bibitem{swain} J.~Swain, Int. J. Mod. Phys. D{\bf 12}, 729 (2003).

\bibitem{nollert} H-P. Nollert, Class. Quant. Grav. {\bf 16}, R159 (1999).

\bibitem{kokkotas} K. D. Kokkotas and B. G. Schmidt, Living Reviews in
Relativity (1999) [gr-qc/9909058].
\bibitem{motl} L. Motl,  Adv. Theor. Math. Phys. {\bf 6},
1135 (2003).
\bibitem{neitzke} L. Motl and A. Neitzke, Adv. Theor. Math. Phys. {\bf 7},
307 (2003).
\bibitem{nollert1} H-P. Nollert, Phys. Rev. {\bf D47}, 5253 (1993).

\bibitem{cardoso} Emanuele Berti, Vitor Cardoso and Shijun Yoshida, gr-qc/0401052;
Emanuele Berti, Vitor Cardoso, Kostas D. Kokkotas and Hisashi Onozawa,
Phys. Rev. D{\bf 68}, 124018 (2003).

\bibitem{hod} S. Hod, Phys. Rev. Lett. {\bf 81}, 4293 (1998).
\bibitem{bek} J. Bekenstein,
in Cosmology and Gravitation, M. Novello, ed. (Atlantisciences, France 2000), pp. 1-85 
[gr-qc/9808028].
\bibitem{steve} Steve Fairhurst, private communication.
\bibitem{km} R.K. Kaul and P. Majumdar, Phys. Rev. Lett. {\bf 84}, 5255 (2000);
Phys. Lett. {\bf B439}, 267 (1998).
\bibitem{gj2} G. Gour and A.J.M. Medved, Class. Quant. Grav. {\bf 20} 3307 (2003);
Phys. Rev. D {\bf 68} 067501 (2003).
\end{thebibliography}
\end{document}